\documentstyle[aps,prd,epsfig]{revtex}
\begin{document}
\renewcommand{\thesection}{\arabic{section}}
\renewcommand{\thesubsection}{\arabic{subsection}}
\title{Collapse/Flattening of Nucleonic Bags in Ultra-Strong Magnetic Field}
\author{Soma Mandal$^{a)}$ {\thanks{E-Mail: soma@klyuniv.ernet.in}}
 and Somenath Chakrabarty$^{a),b)}${\thanks{E-Mail: 
 somenath@klyuniv.ernet.in}} }
\address{
$^{a)}$Department of Physics, University of Kalyani, Kalyani 741 235,
India and
$^{b)}$Inter-University Centre for Astronomy and Astrophysics, Post Bag 4,
Ganeshkhind, Pune 411 007, India
}
\maketitle

\begin{abstract}
It is shown explicitly using MIT bag model that in presence of ultra-strong 
magnetic fields, a nucleon either flattens or collapses in the direction 
transverse to the external magnetic field in the classical or quantum 
mechanical picture respectively. Which gives rise to some kind of mechanical
instability. Alternatively, it is argued that the bag model of 
confinement may not be applicable in this strange situation.
\end{abstract}

\section{Introduction}
One of the oldest subject of physics- "the effect of strong magnetic
field on dense matter" has gotten a new life after the observational
discoveries of a few strongly magnetized exotic stellar objects- known as
magnetars \cite{R1,R2,R3,R4,R5}. These uncommon objects are believed to be 
strongly magnetized young neutron stars and their strong magnetic fields are 
supposed to be the possible sources of X-rays from anomalous X-ray pulsars 
(AXP) and low energy $\gamma$-radiation form the soft gamma-ray repeaters 
(SGR). It is believed that such objects may also act as the central engine 
for gamma ray bursts (GRB). The measured value of magnetic field strength at 
the surface of these objects are $\sim 10^{14}-10^{15}$G. Then it can very 
easily be shown by scalar virial theorem that the magnetic field strength at 
the core region may go up to $10^{18}$G. These objects are also assumed to be 
too young compared to the decay/expulsion time scale of magnetic fields from 
the core region. Now in presence of such intense magnetic fields, most of the 
physical properties of dense stellar matter, e.g., equation of states, 
quark-hadron phase transitions etc., must change significantly \cite{R6,R7,R8}.
Not only that, some of the physical processes \cite{R9,R10}, in particular, 
weak and electromagnetic decays and reactions, neutrino opacities etc., at the 
core region of compact neutron stars will also be affected in
presence of ultra-strong magnetic fields. The transport properties (e.g,
shear and bulk viscosities, thermal and electrical conductivities) of
dense neutron star matter also change both qualitatively and
quantitatively in presence of strong magnetic field \cite{R11,R12}. Furthermore,
these intense magnetic fields could cause structural deformation of the exotic 
objects. In the classical general relativistic theory, it is shown by using 
Maxwell stress tensor that such exotic objects get flattened \cite{R13,R14,R15}
for the macroscopic field $B_m$, whereas in the quantum mechanical scenario 
they collapse in the direction transverse to the magnetic field \cite{R16,R17}.
In the case of ultra-strong magnetic field, the structure of these objects 
could become either disk like (in classical picture) or cigar like (in quantum 
mechanical scenario) from their usual spherical shapes. In the extreme case, 
they may be converted to black disks or black strings. Therefore, in some sense
these strange stellar objects become mechanically unstable in presence of 
ultra-strong magnetic field. Long ago Chandrasekhar and Fermi in their studies 
on the  stability of magnetized white dwarfs explained the possibility of such 
strange behavior \cite{R18}. Those conclusions are also valid for strongly 
magnetized neutron stars, where the white dwarf parameters have to be replaced 
by typical neutron star parameters; the upper limit of magnetic field strength 
for a stable neutron star of typical character is found to be $10^{18}$G. 
In a recent work we have shown that if the 
magnetic field is extremely high to populate only the zeroth Landau level 
(with fully polarized spin states) of electrons, then stable neutron 
star/proto-neutron star matter can not exist in the $\beta$-equilibrium 
condition \cite{R19,R20}. It was also shown by Bander and Rubinstein in the 
context of stability of neutron and protons in a strong magnetic field that in 
presence of extremely strong magnetic field, protons becomes unstable by 
gaining effective mass, whereas neutrons, loosing effective mass and becomes 
stable \cite{R21}. In their calculations a delicate interplay between the 
anomalous magnetic moments of neutron and proton makes the neutron stable and 
proton becomes unstable; decays into neutron via $e^+$ and neutrino emission.

In this article following the recent work of Mart\'{i}nez et al \cite{R16,R17} 
and Kohri et al \cite{R211}, we shall show that even the nucleonic (proton or 
neutron) bags can not be stable in presence of ultra-strong magnetic field- 
they either collapse or elongated in the transverse direction of ultra-strong 
external magnetic field. We have shown that either the nucleons are 
mechanically unstable or the bag model calculations can not be well suited for 
the conditions referred to above. In this work we have therefore studied the 
mechanical stability of a 
neutron/proton placed in an ultra-strong magnetic field. On the other
hand in ref.  \cite{R21}, Bander and Rubinstein have studied the stability of 
these objects from the effective mass point of view and showed that neutrons 
are much more stable energetically than protons in this situation. 
The paper is organized in the following manner: in section 2, we have
reviewed very briefly the MIT bag Lagrangian approach of color
confinement. In section 3, we have studied the collapse of nucleons in
the transverse direction following the ideas of Mart\'{i}nez et al
\cite{R16,R17}. In section 4, following the model proposed by Kohri et
al \cite{R211} in the context of anisotropic $e^+e^-$ pressure, we have
shown that nucleons get flattened in the transverse direction. The
conclusions and discussions are presented in the last section. 
\section{Color Confinement- a Brief Overview}
To study 
the mechanical stability of neutron/proton bags in presence of ultra-strong 
magnetic fields, in the flat space time coordinate, we have considered the MIT 
bag model of quark confinement \cite{R22,R23,R24}.  We have taken into account 
both the gluonic interaction of quarks and the bag pressure $B$ to confine 
quarks within the bag. Before we go into the detailed discussion on the 
mechanical instability problem of nucleonic bags in presence of intense 
magnetic fields, we give a brief overview of bag model Lagrangian approach to 
obtain the pressure balance at the surface of the nucleons. The usual form of 
bag Lagrangian density is given by
\begin{eqnarray}
{\cal{L}}_{{\rm{MIT}}}=[i\{\bar\psi\gamma^\mu\partial_\mu \psi
-(\partial_\mu\bar \psi)\gamma^\mu \psi\}+g\bar \psi
\frac{\lambda_a}{2}\gamma^\mu V_\mu^a \psi &-&
\bar\psi m\psi -\frac{1}{4}F_{\mu\nu}^a F^{\mu\nu a} \nonumber 
-B] \theta_v(x) \nonumber \\ &-&\frac{1}{2}\bar\psi\psi\Delta_s
\end{eqnarray}
where $g$ is the strong coupling constant, $\lambda_a$'s are the $SU(3)$
generators, with $a=1,2,...8$, the gluonic color index, ${V_\mu}^a$ is the 
gluonic field four vector, $F_{\mu\nu}^a$ is the corresponding field tensor, 
$m$ is the current mass of quarks, $B$ is the bag constant, $\theta_v=1$ 
inside the bag and $=0$ outside the bag, $\partial\theta_v/\partial x^\mu=
n_\mu\Delta_s$, $\Delta_s$ is the surface delta-function and  $n_\mu$ is the 
space-like unit vector normal to the surface, The sum over flavors and color 
quantum numbers carried by quarks have not been shown explicitly. To obtain 
the pressure balance at the bag surface, we consider the energy momentum 
tensor of the bag, given by
\begin{eqnarray}
T^{\mu\nu}& = &-g^{\mu\nu}{\cal{L}}+\left(\frac{\partial{\cal{L}}}
{\partial(\partial_\mu\psi)}{\partial^\nu\psi}+\partial^\nu{\bar{\psi}}
\frac{\partial{\cal{L}}}{\partial(\partial_\mu{\bar{\psi}})}\right)
\nonumber\\ & = &-g^{\mu\nu}{\cal{L}}+\frac{i}{2}\left(\bar{\psi}
\gamma^\mu\partial^\nu\psi-(\partial^\nu\bar{\psi})\gamma^\mu\psi\right)
\theta_v
\end{eqnarray}
and using the energy momentum conservation, given by $\partial_\mu T^{\mu\nu}
=0$, we have
\begin{equation}
B\Delta_s n^\nu+\frac{i}{2}\left(\bar{\psi}\gamma^\mu
\partial^\nu\psi-(\partial^\nu\bar{\psi})\gamma^\mu\psi\right) 
n_\mu\Delta_s=0
\end{equation}
and
\begin{equation}
\partial_\mu\left(\bar{\psi}\psi\Delta_s\right)=0
\end{equation}
Now considering the surface boundary condition, given by (obtained from
standard Euler-Lagrange equation)
\begin{equation}
in_\mu\gamma^\mu\psi=\psi
\end{equation}
we obtain on the bag surface
\begin{equation}
B n^{\mu}=\frac{1}{2}\frac{\partial}{\partial{x_\mu}}(\bar{\psi}\psi)
\end{equation}
This equation is nothing but the pressure balance equation.
Since $n^\mu n_\mu= -1$, we have on the bag boundary
\begin{equation}
B=-\frac{1}{2} n_\mu \partial^\mu (\bar{\psi}\psi)
\end{equation}
In the case of spherical bag, $n^\mu\equiv(0,\hat{r})$ and this pressure
balance equation reduces to
\begin{equation}
B=-\frac{1}{2} \frac{\partial}{\partial{r}}(\bar{\psi}\psi)
\end{equation}
Which means that outward pressure of the quarks is exactly balanced by the 
inward vacuum pressure B on the surface of the bag.

\section{Collapse of Nucleonic Bags}
Now we shall consider the nucleonic bag (either neutron or proton) as an interacting thermodynamic system 
in equilibrium. The constituents are valance quarks, sea quarks and gluons.
Then the total kinetic pressure of the system is given by
\begin{equation}
P_{in}=P_{in}^{(v)}+P_{in}^{(s)}+P_{in}^{(g)}
\end{equation}
where $v$, $s$ and $g$ represent the valance quarks, sea quarks and gluonic 
contributions respectively. As discussed before, this internal kinetic 
pressure has to be balanced by the external bag pressure to maintain the 
stability of the system. Then we can write down the effective thermodynamic 
potential per unit volume of the system as
\begin{equation}
-\Omega = P_{in}-B
\end{equation}
and it should be zero. Then following Mart\'{i}nez et al \cite{R16,R17}, 
we have in presence of ultra-strong magnetic field of strength $B_m$, the 
thermodynamic potential per unit volume
(we have chosen 
the gauge $A^\mu\equiv (0,-yB_m/2, xB_m/2,0)$, so that $B_m$ is a constant 
magnetic field along $Z$-axis)
\begin{eqnarray}
T_\mu^\nu&=& \left ( T\frac{\partial \Omega}{\partial T} +\sum_r \mu_r
\frac{\partial \Omega}{\partial
\mu_r}\right )g_\mu^4 g^\nu_4\nonumber \\ &+&4F_{\mu \lambda}F^{\nu
\lambda} \frac{\partial \Omega}{\partial F^2}-g_\mu^\nu \Omega
\end{eqnarray}
Hence the longitudinal component of pressure (along the direction of
field) is given by
\begin{equation}
T_{zz}=P_{\mid \mid}=-\Omega=0
\end{equation}
and the transverse part of total pressure 
\begin{equation}
T_{xx}=T_{yy}=P_\perp =-\Omega-{\cal{M}}B_m= P_{\mid \mid}-{\cal{M}}B_m
\end{equation}
where ${\cal{M}}$ is the effective magnetic dipole moment density of the bag.
 Since $\Omega=0$,
nucleons will therefore be inflated or collapsed
in the transverse direction in presence of ultra-strong magnetic field
depending on the overall sign of ${\cal{M}}$. The system will collapse
if ${\cal{M}}$ is positive, else it will be inflated in the transverse
direction.  In order to have an order of magnitude estimate of extra 
in/out-ward pressure, we choose the contribution to ${\cal{M}}$  
from valance quarks only 
(in fact the valance quarks only contribute in the evaluation of magnetic 
dipole moment of the nucleons). The magnetic dipole moment density of the 
$i$th. component ($i=u$ or $d$-quarks) is given by
\begin{equation}
{\cal{M}}_i=-\frac{\partial \Omega_i}{\partial B_m}
\end{equation}
and the total value is given by 
\begin{equation}
{\cal{M}}=\sum_{i=u,d}{\cal{M}}_i
\end{equation}
where
\begin{equation}
\Omega_i=\frac{g_iq_iB_m}{4\pi^2}\sum_{\nu=0}^{\nu_{\rm{max}}}
\sum_{s=\pm 1} \left [ \mu_i (\mu_i^2-M_{i,\nu,s}^2)^{1/2} -
M_{i,\nu,s}^2 \ln\left (\frac{\mu_i+(\mu_i^2-M_{i,\nu,s}^2)^{1/2}} 
{M_{i,\nu,s}} \right ) \right ]
\end{equation}
is the thermodynamic potential density of the component $i$, $g_i$ and
$q_i$ are respectively the degeneracy and charge of the $i$th. species,
$M_{i,\nu,s}^2=\{(p_\perp^2+m_i^2)^{1/2}+sQ_iB_m\}^2$, $m_i$ is the
current quark mass ($=5$MeV), $p_\perp=(2\nu q_iB_m)^{1/2}$ is the
transverse component of momentum and $Q_i$ is the anomalous
magnetic dipole moment of the $i$th. quark species ($Q_u=1.852\mu_N$ and
$Q_d=-0.972\mu_N$, $\mu_N$ is the nuclear magneton).
The maximum value of Landau quantum number is given by 
\begin{equation}
\nu_{\rm{max}}^{(i)}=\left [ \frac{(\mu_i^2-sQ_iB_m)^2-m_i^2}{2q_iB_m}
\right ]
\end{equation}
where $[~]$ indicates an integer less than the decimal number within the
brackets. To obtain the chemical potentials for $u$ and $d$ quarks, we
have made the following assumptions. The $i$th. quark species density
within the nucleon is given by
\begin{equation}
n_i=\frac{g_iq_iB_m}{2\pi^2}\sum_{\nu=0}^{\nu_{\rm{max}}} \sum_{s=\pm 1}
(\mu_i^2-M_{i,\nu,s}^2)^{1/2} =\frac{{\rm{No.}}(i)}{V}
\end{equation}
where NO($i$) is the number of $i$th. quarks species in the system. Therefore,
NO($i$)=NO($u$)=$1$ for neutrons and $2$ for protons. Similarly,
NO($i$)=NO($d$)=$2$ for neutrons and $1$ for protons and $V$ is the
nucleonic volume. We further assume that $r=0.8$fm
as the radius of the nucleons. Solving numerically, we have obtained the chemical
potentials $\mu_i$'s for both $u$ and $d$ quarks and hence
evaluated the magnetic dipole moment per unit volume
for the system from eqns.(13)-(15). In fig.(1) we have plotted ${\cal{M}}B_m$
for various values of $B_m$ for both neutrons and protons. The product ${\cal{M}
}B_m$ is always positive and oscillatory in the strong field regime ($\geq
10^{17}$G). The system will therefore collapse
in the transverse direction and becomes ellipsoidal with cylindrical symmetry. 
The minor axes lengths $b$ will therefore
oscillate with the strength of magnetic field in particular, above 
$10^{17}$G.  Now in the study of mechanical stability of strongly
magnetized neutron stars in quantum mechanical scenario, it has been
shown that the system will either be inflated or collapsed if the magnetic
dipole moment is negative or positive respectively. It has further been
shown that neutron matter always behaves like a paramagnetic material
with ${\cal{M}}>0$, as a result, in the quantum mechanical picture a
strongly magnetized neutron star always collapses in the transverse direction. 
Therefore we can infer that the conclusion drawn for such macroscopic objects 
like neutron stars is also valid in the microscopic level- e.g., neutrons or 
protons. We can then conclude that in a strong magnetic field, not only 
neutron stars, even their constituents, neutrons and protons become 
mechanically unstable. Alternatively, one could conclude that the bag model is 
perhaps not applicable in such strange situation, in that case the use of bag 
model for magnetized quark stars is also questionable. Therefore the
investigations of this section show that both neutrons and protons
become cigar like and in the extreme case they may reduce to what is
called black string.

\section{Flattening of Nucleons}

In this section we shall evaluate the longitudinal and transverse parts
of the kinetic pressures following ref.\cite{R211}. We choose the
gauge $A^\mu\equiv (0,0,xB_m,0)$ so that $\vec B_m\equiv (0,0,B_m)$.
Then the solution of the Dirac equation is given by
\begin{equation}
\psi=\exp(-iE_nt)
\left (\begin{array}{c}
\phi \\ \chi 
\end{array} \right )
\end{equation}
where $\phi$ and $\chi$ are the upper and lower components. The upper
component is given by
\begin{equation}
\phi=\exp(ip_yy+ip_zz)f_n\zeta_s,
\end{equation}
where $n=0,1,2,..$ is the Landau quantum number, $s=\pm1$, the spin
quantum number, so that 
\begin{equation}
\zeta_1=\left (\begin{array}{c}1\\0 \end{array} \right ),
\end{equation}
\begin{equation}
\zeta_{-1}=\left ( \begin{array}{c}0\\1 \end{array} \right )
\end{equation}
and
\begin{equation}
f_n(x,p_y)=\frac{1}{(2^nn! \pi^{1/2})^{1/2}}\exp\left (
-\frac{\xi^2}{2} \right )H_n(\xi)
\end{equation}
$\xi=(q_iB_m)^{1/2}(x-p_y/(q_iB_m))$ and $H_n(\xi)$ is the Hermite
polynomial of order $n$. The lower component is given by
\begin{equation}
\chi=\frac{\vec \sigma . (\vec p-q_i \vec A)}{E_n+m_i}\phi
\end{equation}
The energy eigen value is given by
$E_n=(p_z^2+m_i^2+q_iB_m(2n+1-s))^{1/2}$. Then we have from the first
part of eqn.(2), 
\begin{equation}
T^\mu_\nu={\rm{diag}}(E_n,-\hat P_x, -\hat P_y, -\hat P_z)
\end{equation}
whereas all the off-diagonal terms are zero. Then it is very easy to
show
\begin{equation}
\hat P_x=\hat P_y=\left (n+\frac{1}{2}-\frac{s}{2} \right
)\frac{q_iB_m}{E_n}, \hat P_z=\frac{p_z^2}{E_n}
\end{equation}
These are called the dynamic pressure \cite{R211}. The ensemble average
of these pressures are given by (at $T=0$)
\begin{equation}
P_x=P_y=\frac{(q_iB_m)^2}{2\pi^2}\sum_{\nu=0}^{[\nu_{\rm{max}}]} \nu (2-
\delta_{0\nu}) 
\ln \left \{
\frac{\mu_i+(\mu_i^2-m_\nu^2)^{1/2}}{m_\nu}\right \}
\end{equation}
which are the transverse part,
where $2\nu=2n+1-s$ and $m_\nu^2=m_i^2+2q_iB_m\nu$. Similarly, we have the
longitudinal component
\begin{equation}
P_z=\frac{q_iB_m}{4\pi^2}\sum_{\nu=0}^{[\nu_{\rm{max}}]}(2-
\delta_{0\nu}) \left [ \mu_i(\mu_i^2-m_\nu^2)^{1/2}-
\ln \left \{
\frac{\mu_i+(\mu_i^2-m_\nu^2)^{1/2}}{m_\nu}\right \} \right ]
\end{equation}
Following the same numerical techniques as followed in previous section,
in figs.(2) and (3) we have plotted the longitudinal and transverse component 
of kinetic pressures
with various magnetic field strengths for protons and neutrons
respectively. The curves in both the figures show that the
longitudinal part of kinetic pressure is zero and /  or very low for
high magnetic field strength. Whereas the transverse part is high for
high magnetic field. These two components saturate to some constant value 
for low or
moderate magnetic field strength. Which indicates that the system reduces to
pressure isotropic configuration at low magnetic field (as we generally
see in conventional thermodynamic system). Therefore, according to 
this model, at very high magnetic field strength the
system (neutron or proton) becomes oblet in shape and in the extreme 
case it reduces to a black disk.  
\section{Conclusions}
In conclusion, we have studied the mechanical stability of neutrons and protons
in a compact neutron star in presence of strong quantizing magnetic field. We
have followed two entirely different approaches. In the so called quantum
mechanical picture, in which the interaction of magnetic dipole moments
of quark constituents with the external magnetic field has been
considered, the shapes of both neutron and proton become prolate type from their
usual spherical nature. The effect is more
prominent at high field limit ($ > 10^{16}$G). On the other hand, in the
classical picture, both the systems acquire oblate shape. The effect is
again prominent for high magnetic field. In the classical picture, it has
been observed that such anisotropy of kinetic pressure is automatically removed 
at
moderate ($\geq 10^{15-16}$G) values of magnetic field strength and both the
systems become mechanically stable. However, in the quantum mechanical
picture, there is always an extra in-word pressure in the transverse
direction even for moderate values of magnetic field strength. This is
because of non-zero finite values for ${\cal{M}}B_m$ of the systems, but
the effect is not so significant. Therefore, the behavior of bulk
objects like neutron stars and their constituents, e.g., neutrons and
protons (which are of microscopic in sizes) are almost identical in an 
external strong magnetic field.
\begin{figure} 
\psfig{figure=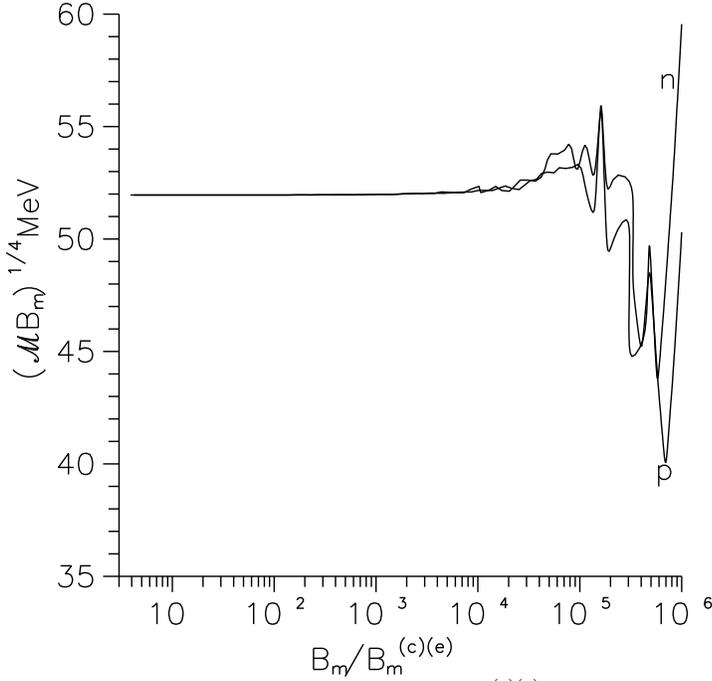,height=0.5\linewidth}
\caption{ The variation of ${\cal{M}}B_m$ with $B_m/B_m^{(c)(e)}$ for
neutron (indicated by the symbol $n$) and proton (indicated by the
symbol $p$).}
\end{figure}
\begin{figure} 
\psfig{figure=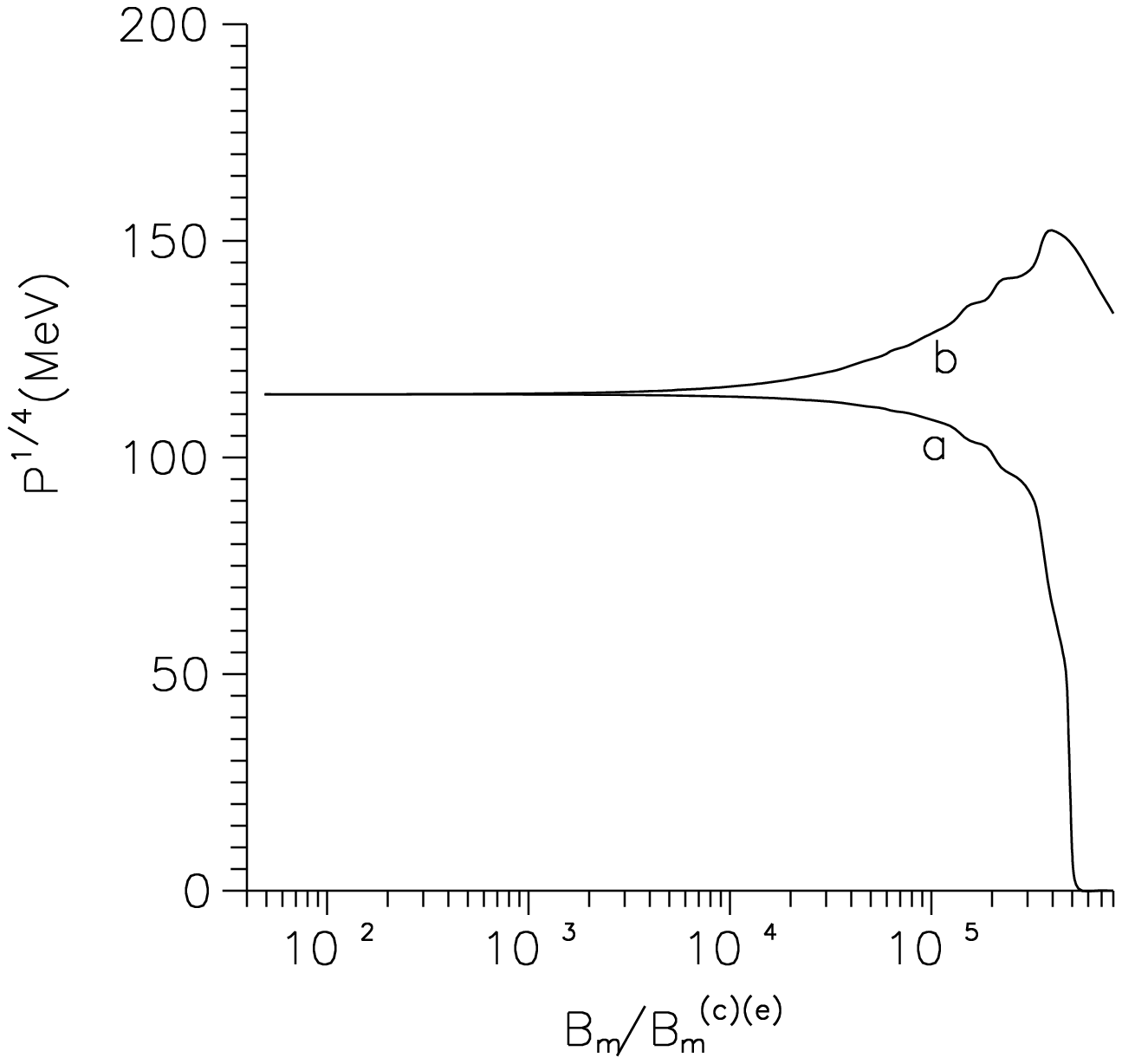,height=0.5\linewidth}
\caption{ The variation of kinetic pressure ($P^{1/4}$ in MeV) with 
$B_m/B_m^{(c)(e)}$ for
proton. Curve $a$ is for longitudinal and $b$ is for transverse
components of kinetic pressures respectively.  }
\end{figure}
\begin{figure} 
\psfig{figure=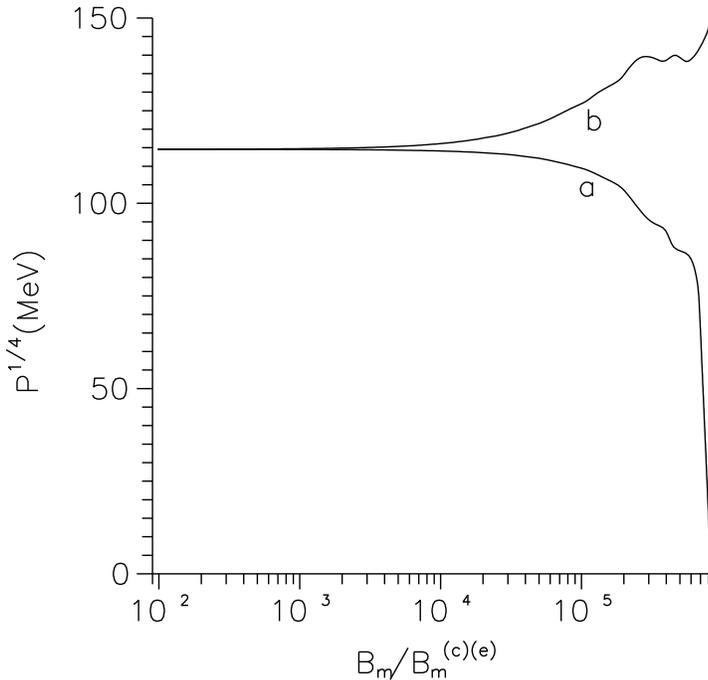,height=0.5\linewidth}
\caption{ The variation of kinetic pressure ($P^{1/4}$ in MeV) with 
$B_m/B_m^{(c)(e)}$ for
neutron. Curve $a$ is for longitudinal and $b$ is for transverse
components of kinetic pressures respectively.  }
\end{figure}

\noindent {\sl{Acknowledgment: SC is thankful to the Department of Science and 
Technology, Govt. of India, for partial support of this work, Sanction 
number:SP/S2/K3/97(PRU).}}
\end{document}